\documentclass[conference]{IEEEtran}
\usepackage{cite}
\usepackage{amssymb,amsmath,amsthm,enumitem}
    
    \DeclareMathOperator*{\argmin}{arg\,min}
    \newcommand{\norm}[1]{\left\lVert#1\right\rVert}
\usepackage{tabstackengine} 
\usepackage{algorithmic}
\usepackage{graphicx}
\usepackage{textcomp}
\usepackage{xcolor}

\usepackage[caption=false,font={small}]{subfig}
\usepackage{tabulary}
\usepackage{breqn}
\usepackage{bm}
\usepackage{csquotes}
\usepackage{color}

\usepackage{layouts}
\usepackage[font=small,skip=5pt]{caption} 

\newcommand{\armar}{\IEEEauthorrefmark}

\newcommand\blfootnote[1]{%
    \begingroup
    \renewcommand\thefootnote{}\footnote{#1}%
    \addtocounter{footnote}{-1}%
    \endgroup
}

\begin{document}	


\title{GPU Acceleration for Synthetic Aperture Sonar Image Reconstruction}

\makeatletter
\patchcmd{\@maketitle}
  {\addvspace{0.5\baselineskip}\egroup}
  {\addvspace{-1\baselineskip}\egroup}
  {}
  {}
\makeatother

\makeatletter
\newcommand{\linebreakand}{%
\end{@IEEEauthorhalign}
\hfill\mbox{}\par
\mbox{}\hfill\begin{@IEEEauthorhalign}
}
\makeatother
\author{
	\IEEEauthorblockN{Isaac D. Gerg\armar{1}, Daniel C. Brown\armar{1}, Stephen G. Wagner\armar{1}, Daniel Cook\armar{2}, Brian N. O'Donnell\armar{2}, \\Thomas Benson\armar{3}$^{1}$, Thomas C. Montgomery\armar{1}}\hspace{0.3cm}
	\linebreakand
	\IEEEauthorblockA{\textit{\armar{1}Applied Research Laboratory} \\
		\textit{Pennsylvania State University}\\
		State College, PA USA}
	\and
	\IEEEauthorblockA{\textit{\armar{2}Georgia Tech Research Institute} \\
		Atlanta, GA USA}
  	\and
    \IEEEauthorblockA{\textit{\armar{3}Smiths Digital Forge} \\
        Fremont, CA USA}
}
\maketitle
\thispagestyle{plain}
\pagestyle{plain}

\begin{abstract}
Synthetic aperture sonar (SAS) image reconstruction, or beamforming as it is often referred to within the SAS community, comprises a class of computationally intensive algorithms for creating coherent high-resolution imagery from successive spatially varying sonar pings. Image reconstruction is usually performed topside because of the large compute burden necessitated by the procedure.  Historically, image reconstruction required significant assumptions in order to produce real-time imagery within an unmanned underwater vehicle's (UUV's) size, weight, and power (SWaP) constraints.  However, these assumptions result in reduced image quality.  In this work, we describe ASASIN, the Advanced Synthetic Aperture Sonar Imagining eNgine. ASASIN is a time domain backprojection image reconstruction suite utilizing graphics processing units (GPUs) allowing real-time operation on UUVs without sacrificing image quality.  We describe several speedups employed in ASASIN allowing us to achieve this objective. Furthermore, ASASIN's signal processing chain is capable of producing 2D and 3D SAS imagery as we will demonstrate.  Finally, we measure ASASIN's performance on a variety of GPUs and create a model capable of predicting performance. We demonstrate our model's usefulness in predicting run-time performance on desktop and embedded GPU hardware.

\end{abstract}

\vspace{-1\baselineskip}

\blfootnote{$^{1}$Work performed while at the Georgia Tech Research Institute.}

\section{Brief Introduction to Synthetic Aperture Sonar}

In echo-based imaging techniques, aperture (or antenna) size is inversely proportional to image resolution.  However, there are practical limits to the physical aperture size.  From a historical perspective, the radar community addressed these physical limits first and overcame through the advent of synthetic aperture processing.  In this technique, the platform transmits while moving and the aperture is synthetically formed by coherently combining the reflected signals over several transmissions.  This results in a synthetically lengthened aperture which is much larger than the physical aperture resulting in improved resolution. This type of processing forms basis of what we understand today as synthetic aperture radar (SAR). 

Synthetic aperture sonar (SAS) image formation techniques grew out of traditional SAR methods. However, there are two key facets between SAS and SAR that play a role in the algorithmic differences:
\begin{enumerate}
    \item Significant differences in the speed of the medium (i.e. $1.5\times10^{3}$ m/s versus $3\times10^{8}$ m/s). This often has to be accounted for in SAS but can be generally ignored in SAR.
    \item Platform motion between transmissions can be quite severe in underwater environments.  With no global position system (GPS) reference available, platform motion must be estimated from the collected sonar data. Inertial measurement units (IMUs) embedded in UUVs do not provide sufficient accuracy to form the synthetic aperture \cite{bellettini2002theoretical}. It's worth mentioning though that high precision satellite positioning has been used with surface-based SAS systems (i.e. downward looking, surface ship mounted sonars) to provide sufficient positional accuracy \cite{brown2019simulation}.
\end{enumerate}

Operational image formation techniques for SAS and SAR have been developed with computational efficiency in mind since the the time domain backprojection algorithm, simple to understand and capable of producing quality imagery in high-motion environments, can be quite expensive to compute. Many image reconstruction systems implement either $\omega$-k or fast factorized backprojection based methods due to their computational efficiency but often at the expense of numerical approximations resulting in reduced image quality in some environments.  With the advent of GPUs, time domain backprojection algorithms can now operate in real-time \cite{baralli2013gpu}. For those looking for an in-depth description of challenges specific to SAS image reconstruction, \cite{callow} and \cite{cook2007synthetic} provide good overviews.

This work describes a GPU-accelerated time domain backprojection image reconstruction engine we call ASASIN. ASASIN is capable of creating 2D and 3D SAS imagery.
\textbf{Specifically, this work provides the following technical contributions:}
\begin{enumerate}
    \item We describe a processing chain capable of producing high-fidelity 2D and 3D SAS imagery using a time domain backprojection algorithm and outline speedups used to improve the run-time performance.
    \item We propose a robust platform motion estimation algorithm capable of generating high-quality imagery even when severe ping-to-ping motion is present.
    \item We demonstrate ASASIN's ability to run real-time embedded in a UUV by benchmarking it on a variety of contemporary GPUs including two embedded models.
\end{enumerate}
The sections following provide an overview of the time domain backprojection method for image reconstruction, the ASASIN implementation of our image reconstruction pipeline for efficient computation on GPUs, and experimental results comparing ASASIN compute performance on a variety of commodity GPU hardware.

\section{Proposed Method: ASASIN}

ASASIN consists of a framework that executes a modular sequence of algorithms for SAS image reconstruction.  The input to ASASIN is a configuration file containing all of the processing parameters and time-series hydrophone array recordings. The output is image data products, debug, and processing logs. \figurename \ref{high_level_diagram_asasin} shows this schematically.  The configuration file guarantees reproducibility in the processing as well as the ability to quickly compare results from different parameter settings.  Input hydrophone time-series data can be in a variety of formats as long as an appropriate reader is present which is dictated by the configuration file.  All data readers are derived from a base class and implement the necessary methods to read data and navigation information.  ASASIN outputs several types of data products including GeoTiff (multi-layer GeoTiff for the 3D imaging case).

\begin{figure}[t]
    \includegraphics[width=0.75\linewidth]{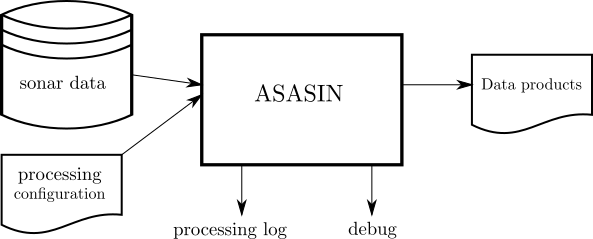}
    \centering
    \caption{High-level diagram of the ASASIN processor.  ASASIN ingests the raw sonar data along with a configuration file containing the processing parameters.  During processing, ASASIN outputs a processing log containing performance metrics, processing path choices, and data issues.  Also output are several debug products used to assess the processing quality.  At the end of processing, the final output is usually an image data product like GeoTIFF.}
    \label{high_level_diagram_asasin}
\end{figure}

\subsection{Time Domain Backprojection Algorithm}
Image reconstruction with a backprojection algorithm requires inverting the acoustic signals received by the sonar to estimate the seafloor acoustic reflectivity.  The time-series that is inverted is modeled as Eq. \ref{eqn:backprojection} and depicted for a single pixel in \figurename \ref{fig:scene_forward_model}. The backprojection algorithm seeks to find $\sigma(\mathbf{x})$ from $e(t, \mathbf{x}_{RX}(t))$ through inversion.  Often the sonar system is not calibrated so estimating $\sigma (\cdot)$ exactly is not possible. However, one can still form a suitable image by making very simplistic assumptions on the form of $\sigma(\cdot)$.

Any pair of pixel locations in Eq. \ref{eqn:backprojection} can be computed independently allowing the operation to be to be highly parallelizable, and thus is suitable for GPU acceleration.

\begin{table*}
    \centering
    \begin{minipage}{\textwidth}
        \begin{equation}
            e(t,\mathbf{x}_{RX}(t)) \approx \int \frac{\sigma(\mathbf{x})}{\norm{\mathbf{x}_{TX}-\mathbf{x}} \norm{\mathbf{x}_{RX}(t)-\mathbf{x}}} \cdot q\left(t - \frac{\norm{\mathbf{x}_{TX}-\mathbf{x}} + \norm{\mathbf{x}_{RX}(t)-\mathbf{x}} }{c}\right) d\mathbf{x}
            \label{eqn:backprojection}
        \end{equation}
       where $e(t, \mathbf{x}_{RX}(t))$ represents the time-series of the array, $t$ is time, $\mathbf{x}_{TX}$ is the transmitter location (we assume it's stationary during transmit), $\mathbf{x}_{RX}(t)$ is the receiver location, $\mathbf{x}$ is the scattering location, $\sigma$ is the acoustic scattering cross-section function \cite{hunter2003simulation}, $c$ is the speed of sound, and $q$ is the transmitted signal waveform. \figurename \ref{fig:scene_forward_model} gives a depiction of this process for one pixel of the integration.
        \medskip
        \hrule
    \end{minipage}
\end{table*}

\begin{figure}[t]
    \includegraphics[width=0.95\linewidth]{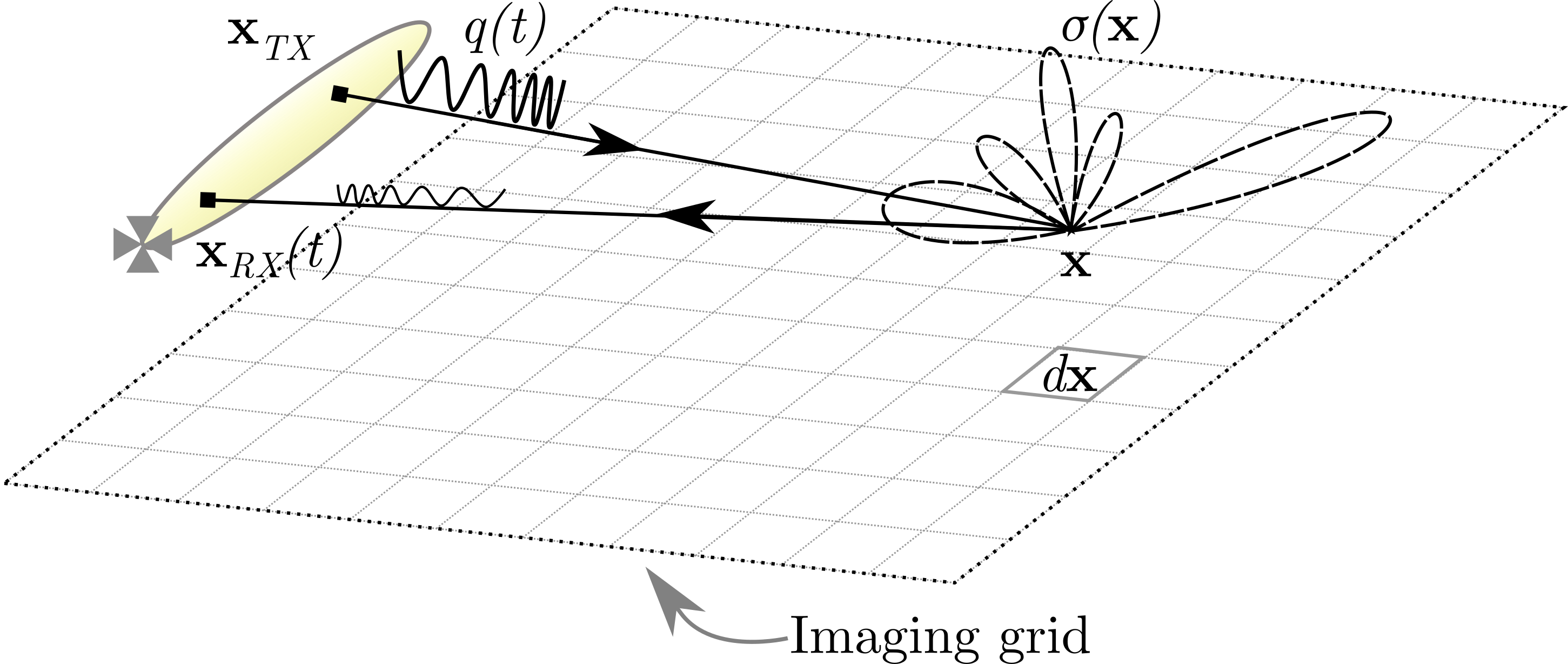}
    \centering
    \caption{Depiction of the forward model described by Eq. \ref{eqn:backprojection}. ASASIN uses a time domain backprojection algorithm for image reconstruction where the forward model is inverted to solve for $\sigma(\mathbf{x})$ for each pixel.  Since each pixel in the integration can be processed independently, this formulation lends itself to being straightforward to parallelize on a GPU.}
    \label{fig:scene_forward_model}
\end{figure}

\subsection{Coordinate System Description of ASASIN}
Image reconstruction is formally a navigation problem linking several coordinate systems. ASASIN uses a north-east-down (NED) coordinate system noted by $x, y, \text{and } z$ respectively shown in \figurename \ref{fig:scene_geometry}. Transducer elevation is noted as $\phi$ representing the full-width-half-maximum power (FWHM) beam.  Azimuthal beamwidth is denoted by $\theta$, also FWHM. Attitude manipulation of points in this space is given by the transformation of Eq. \ref{eqn:rollpitchyaw}. The imaging grid for a stripmap collection is shown in \figurename \ref{fig:scene_geometry}. Notice that the first ping and last ping positions are outside the imaging grid.  This is done intentionally so that the far range corners of the image have full angular support from the beam during the backprojection process making the image resolution constant throughout the scene.

We place the origin of the scene directly below the vehicle and on the seafloor of the very first ping position.  This convention yields two noteworthy attributes:
\begin{enumerate}
    \item The position of the first ping is directly above the origin and thus has a negative $z$ position due to the definition of the origin and our NED coordinate system. (The sonar platform operates at -$z$ altitude.)
    \item The start of the imaging grid is not that origin but an x-position which is a function of the sensor beamwidth as shown in \figurename \ref{fig:top_down}.
\end{enumerate}

\begin{table*}
    \centering
    \begin{minipage}{\textwidth}
        \begin{equation}
            \boldsymbol{\widetilde{p}} = \begin{bmatrix}
            \cos(\theta) \cos(\psi) & \cos(\psi) \sin(\theta) \sin(\phi) - \cos(\phi) \sin(\psi) & \cos(\phi) \cos(\psi) \sin(\theta) + \sin(\phi)sin(\psi) \\
            \cos(\theta) \sin(\psi) & \cos(\phi) \cos(\psi) + \sin(\theta) \sin(\phi) \sin(\psi) & \cos(\phi) \sin(\theta) \sin(\psi) - \cos(\psi)\sin(\phi) \\
            -\sin(\theta) & \cos(\theta) \sin(\phi) & \cos(\theta) \cos(\phi)
            \end{bmatrix}
            \boldsymbol{p}
            \label{eqn:rollpitchyaw}
        \end{equation}
        where $\phi$, $\theta$, and $\psi$ are the platform roll, pitch, and yaw respectively in radians. We use the following conventions of angles: positive roll lowers  the starboard side, positive pitch raises the bow, and positive yaw rotates to starboard. 
        \medskip
        \hrule
    \end{minipage}
\end{table*}

\begin{figure}[t]
    \includegraphics[width=0.95\linewidth]{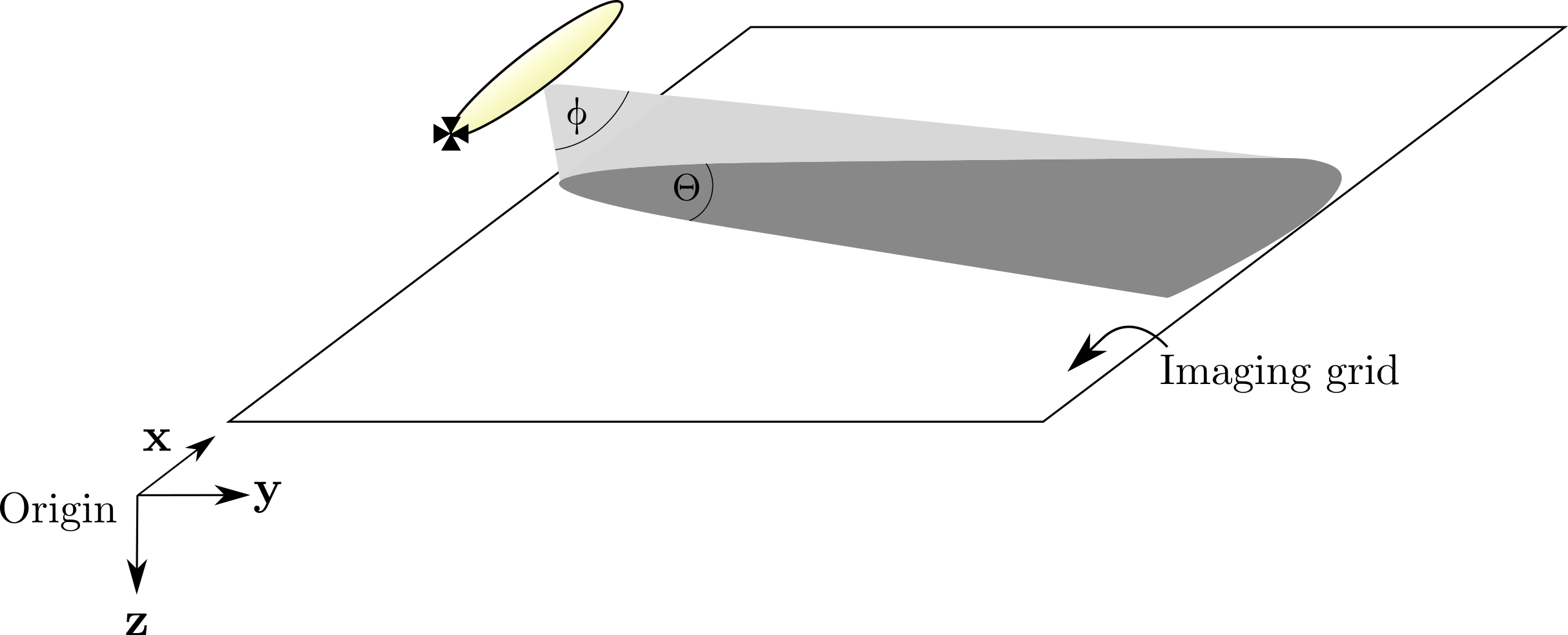}
    \centering
    \caption{A well-defined imaging geometry aids in data and debug output interpretation. Shown here is the imaging geometry and coordinate system used in ASASIN for stripmap imaging. The origin is defined outside the imaging grid so the image corners have full angular support to deliver constant resolution throughout the scene. $\phi$ and $\theta$ denote the FWHM elevation and azimuth beamwidths respectively.}
    \label{fig:scene_geometry}
\end{figure}

\begin{figure}[t]
    \includegraphics[width=0.95\linewidth]{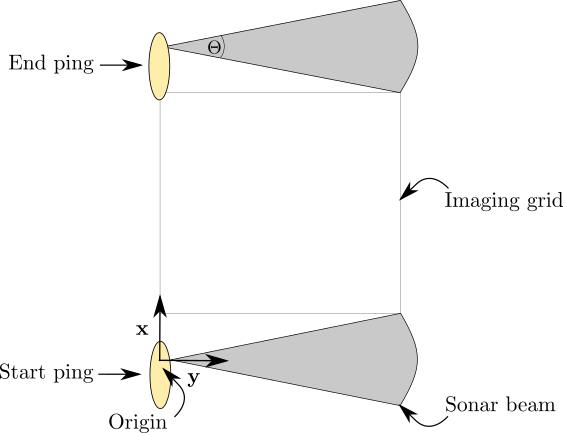}
    \centering
    \caption{Top-down view of the imaging geometry for stripmap imaging.  Placing the origin outside the imaging grid gives the image corners appropriate angular support and consequently uniform image resolution across the scene. $\theta$ denotes the FWHM azimuthal beamwidth.}
    \label{fig:top_down}
\end{figure}

\begin{figure}[t]
    \includegraphics[width=0.85\linewidth]{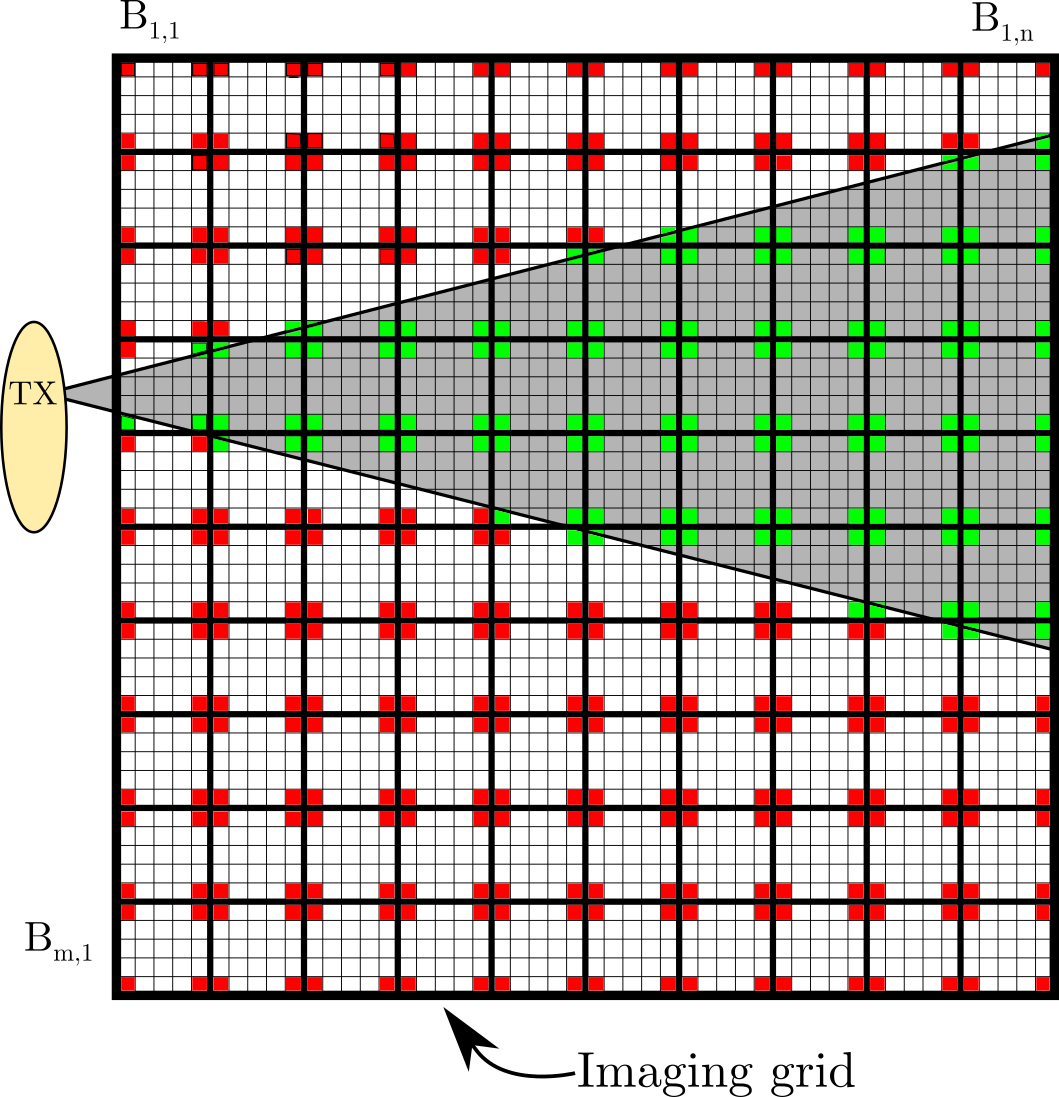}
    \centering
    \caption{Illustration of the ray-culling algorithm used to prune the processing space of the backprojection algorithm. The scene is divided into subblocks where each corner tested to determine if it is in the sonar beam. The red squares indicate the sonar beam is within the FOV of the corner pixel; green indicates otherwise.  Any block containing at least one green pixel is further processed, while blocks containing only red pixels are further ignored.}
    \label{fig:rayculling}
\end{figure}

\subsection{Computational Considerations to Improve Run-Time Performance}
We were able to increase run-time performance over the initial parallelization of Eq. \ref{eqn:backprojection} by exploiting the problem geometry and particulars of the GPU hardware.  This was accomplished through three primary means. First, we implemented our own memory allocator for the GPU. This prevents memory fragmentation with ASASIN and other GPU processes running simultaneously (e.g. a display/GIS tool).  We implemented this by structuring our algorithms to derive from a generic base class which requires each derived class to implement a method returning the amount of memory needed for execution.  One large memory slab is allocated to accommodate all the requests as algorithms are instantiated by ASASIN.   Machine learning practitioners will recall our memory allocation scheme is similar to the Tensorflow library \cite{tensorflow}.

Second, we utilized GPU hardware texture units to perform fast sensor position lookup.  The texture units are capable of performing linear interpolation in hardware so we upsample the sensor positions and store them there.  Upon querying the sensor position for a specific time, the texture units quickly interpolate (in hardware) from our upsampled table and return the position.

Finally, we utilize ray culling to speed up the integration of a single ping.  For each ping in the sonar collection, we must integrate the time-series into the appropriate complex pixels.  This procedure requires every pixel to query the sonar geometry and determine if it is in the sonar field of view (FOV); such a search has compute complexity $O(n^2)$. For the majority of SAS systems, the beamwidth is sufficiently small that each ping insonofies a fraction of the total imaging grid.  To speed up this computation, we take advantage of the fact that the receive and transmit beams are conically shaped and thus convex.  We exploit the convexity by performing a coarse-to-fine block search to quickly prune areas requiring no computation. This is done by first dividing the imaging grid into square blocks of size $b$ pixels.  Next, for each block, $B_{i,j}$, we test its four corners for transmitter/receiver visibility.  Finally, if none of the tests pass, we discard the block for further processing. When any corner of the block passes the test, we further process all the pixels in the block.

\figurename \ref{fig:rayculling} depicts an example of the ray culling procedure for a single ping.  The imaging grid is composed of $m\times n$ blocks each of size 5 pixels x 5 pixels. The gray area shows the FOV of the transmitter and receiver.  The four corners of each block, $B_{i,j}$, are tested to see if they are in the FOV of the sonar.  Pixels that are in the field of view are marked in green and those that are not are marked in red.  Blocks with no green corners are pruned from the FOV search and an explicit FOV test for each pixel is performed for all those remaining. We find the ray culling procedure accelerates our processing by a factor of two.

\subsection{Motion Estimation of Platform Heave, Sway, and Surge}
Inverting Eq. \ref{eqn:backprojection} for $\sigma$ requires precise knowledge of positions $\mathbf{x}$ and $\mathbf{x}^\prime$ for each ping to produce focused imagery.  The inversion is straightforward once these quantities are known. Hence, the motion estimation of the sensor platform deserves much attention and several methods have been developed to estimate it \cite{cook2007synthetic,marston2016volumetric,bellettini2002theoretical}.

ASASIN estimates sensor position by a derivative work of the displaced phase center (DPC) method described in \cite{cook2007synthetic}.  In this work, time-delays are measured from overlapped coherent looks of the seafloor of consecutive pings.  The amount of time-delay measured is a function of the translation of the second ping to the first mainly in the sway and heave directions.  Applying this principle as a function of range can unambiguously determine the sway and heave translation between the two pings.

Four degrees of freedom (surge, roll, pitch, and yaw) remain to be addressed. For surge, ASASIN can select from two sources: Doppler velocity logger (DVL) measurement or the along track estimation (ATE) algorithm of \cite{ATE}.  For vehicle attitude (roll, pitch, and yaw), ASASIN uses the measurements provided by an on-board, high-fidelity inertial navigation system (INS).  

Estimation of the remaining degrees of freedom, platform sway and heave, will be the primary focus of this section. At it's simplest, ASASIN's motion model is described by Eq. \ref{eqn:g} and is composed of three primary parts.  The first part describes the time-of-flight from the first ping transmission, its reflection off the seafloor, and its reception by a receive element.  The second part is identical to the first but now concerning the second ping. Finally, the third part is the measured time-delay between the two pings. This model makes the assumption that sonar transmission occurs instantly and the sonar platform is continually moving during reception.  Forgoing the latter assumption is commonly called the \emph{stop-and-hop} approximation.

In Eq. \ref{eqn:g}, a single time-delay estimate is measured from a common, overlapped patch of seafloor illuminated by consecutive pings.  Assuming the measured time-delay is noiseless, possible platform heave and sway estimates are those which equate $g$ to zero. Thus, for a single time-delay estimate, many solutions exist.  Therefore, several time-delay measurements as a function of range are required to generate at a unique solution.  Such a function ingesting hypothesized parameters of a model and returning a scalar error is referred to as a \emph{residual}. Since the time-delay estimates are noisy, we opt to minimize the sum of all squared residuals, $g^2$, from all the measurements. Assuming $k$ time-delay estimates measured as a function of range / reference time $t_k$ for a given ping pair, $i$, our loss for estimating heave and sway among all ping pairs, which we call $\mathcal{L_{\text{DPC}}}$, is given in Eq. \ref{eqn:motion}. Since we assume continual motion during reception, Eq. \ref{eqn:motion} models the position of the platform as a first order kinematic equation (i.e. $\mathbf{p} = \mathbf{v}t$).  Thus, the minimization reduces to determining the platform velocity (specifically the $v_y$ and $v_z$ components) of each ping through minimization of $g$ for all time-delay measurements for all ping pairs.  We have experimented with higher order kinematic models but found they provide little benefit.  One last item of information is necessary to compute the specific transmitter and receiver locations of each ping in Eq. \ref{eqn:g}: the vehicle attitude and lever arm information which we define as $\mathbf{\Omega}$. This information translates the vehicle position, $\mathbf{p}$, for each ping to the components of $\mathbf{p}_{\text{RX}}$ and $\mathbf{p}_{\text{TX}}$ necessary for computation of $g$.

It is worth explicitly mentioning four important attributes resulting from  Eq. \ref{eqn:motion}.  The first attribute is the linkage between the velocity estimates of all the pings.  The estimate $\mathbf{v}_{i+1}$ depends on the estimates of $\mathbf{v}_{i}$ and $\mathbf{v}_{i+2}$. Therefore, all the ping velocities are estimated jointly.  The second attribute is the ease of which we can weight each residual inside the sum of Eq. \ref{eqn:motion}.  A common weighting scheme is to use the correlation coefficient associated with $\delta_{i,k}[t_k]$ which is proportional to the measurement's signal-to-noise ratio.  The third attribute is the ease of which the residuals can be weighted to mitigate outliers through proper choice of $h$.  In Eq. \ref{eqn:motion}, each residual is squared before accumulation by the sum.  However, squared-loss can be sensitive to outliers and assign them more weight than desired. This can be mitigated by replacing the squared-loss with another loss which de-weights the contribution of large loss terms (i.e. probable outliers) in the sum. This is easily accomplished by replacing $h$ in Eq. \ref{eqn:motion} with the appropriate convex function such as the Huber loss given in Eq. \ref{eqn:huber}. 
\begin{equation}
\label{eqn:huber}
h(a) = \begin{cases}
\frac{1}{2}a^2  & \text{for } \vert a \vert \leq 1 \\
\vert a \vert - \frac{1}{2} & \text{otherwise}
\end{cases}
\end{equation}
Finally, the fourth attribute is that any bathymetry information known about $\mathbf{p}_s$ can be easily integrated into the model as the $z$-component (for a flat seafloor assumption, $\mathbf{p}_{s,z} = 0$).

\begin{table*}
    \centering
    \begin{minipage}{\textwidth}
    \begin{equation}
        \label{eqn:g}
        g(\mathbf{p}_{\text{TX}_1}, \mathbf{p}_{\text{RX}_1}(t), \mathbf{p}_{\text{TX}_2}, \mathbf{p}_{\text{RX}_2}(t), \mathbf{p}_{s}, \delta(t), c) = c^{-1} 
        \bigl( 
        \underbrace{(\| \mathbf{p}_{\text{TX}_1} - \mathbf{p}_{s} \| + \| \mathbf{p}_{s} - 
        \mathbf{p}_{\text{RX}_1}(t)\|)}_{\text{Path of TX1 to Seafloor to RX1}}
        - \underbrace{(\norm{\mathbf{p}_{\text{TX}_2} - \mathbf{p}_{s}} + \| \mathbf{p}_{s} - \mathbf{p}_{\text{RX}_2}(t)\|)}_{\text{Path of TX2 to Seafloor to RX2}} \bigr) + \underbrace{\delta(t)}_{\text{Time-Delay Between Acoustic Returns}}
    \end{equation}
    Our fundamental residual equation for our motion model. $\delta(t)$ is the measured time difference of arrival between ping one and ping two to a common acoustic footprint location on the seafloor, $\mathbf{p}_{s}$, at a reference time $t$, $\mathbf{p}_{\text{TX}_1}$ is the position of ping one during transmit, $\mathbf{p}_{\text{RX}_1}(t)$ is the position of the receive element of ping one after time $t$, $\mathbf{p}_{\text{TX}_2}$ is the position of the ping two during transmit, $\mathbf{p}_{\text{RX}_2}(t)$ is the position of the receive element of ping two after time $t$, and $c$ is the speed of sound in the medium. We make the approximation that the ping transmission occurs instantaneously. 
    \medskip
    \hrule
    \end{minipage}
\end{table*}

\begin{table*}
    \centering
    \begin{minipage}{\textwidth}
    \begin{equation}
    \mathcal{L_{\text{DPC}}} = \sum_{i,k} h(g(\mathbf{p}_{\text{TX}_i},\mathbf{p}_{\text{RX}_i}(t_k), \mathbf{p}_{\text{TX}_{i+1}}, \mathbf{p}_{\text{RX}_{i+1}}(t_k), \mathbf{p}_{s}, \delta_{i,k}[t_k], c \vert \mathbf{v}_{i}, \mathbf{v}_{i+1}, \mathbf{\Omega}_{i}, \mathbf{\Omega}_{i+1}))
    \label{eqn:motion}
    \end{equation}
    The DPC-related loss estimating the $y$- and $z$- components of $\mathbf{v}$.  $\mathbf{\Omega}_i$ contains the associated lever arms and vehicle attitude measurements for ping $i$ needed to compute the transmitter and receiver positions at time $t_k$ using the proposed heave and sway velocity components of $\mathbf{v}_i, \mathbf{v}_{i+1}$. $h$ is the loss function which we define as $h(a) = a^2$ yielding a sum-of-squared errors/residuals expression.
    \medskip
    \hrule
    \end{minipage}
\end{table*}

We add two additional regularization terms to Eq. \ref{eqn:motion} to smooth the solution and have it not drift too far from the Doppler velocity logger (DVL) measurements. Equations \ref{eqn:smooth} and \ref{eqn:dvl} describe these regularization terms respectively,
\begin{equation}
    \mathcal{L_{\text{smooth}}} = \sum_{i}  ( {v_y}''[i] )^2 +   \sum_{i}  ( {v_z}''[i] )^2
    \label{eqn:smooth}
\end{equation}

\begin{equation}
    \mathcal{L_{\text{DVL}}} = \sum_i \left[ \left( p_{z,1}(0) +    \sum_{n=1}^{i}   v_z[n] \Delta_{n} \right)  - z_{\text{DVL}}[i] \right ] ^2
    \label{eqn:dvl}
\end{equation}
where $p_{z,i}(t)$ is platform z-position position of ping $i$ at time $t$ since transmit (e.g. $p_{z,1}(0)$ is the z-position of of first ping at transmission) and $\Delta_{n}$ is the time difference between transmissions of ping $n$ and ping $n+1$. Velocities are converted to positions through integration. The final loss used for optimization is a combination of Eq. \ref{eqn:motion}, Eq. \ref{eqn:smooth}, and Eq. \ref{eqn:dvl} and is given in Eq. \ref{eqn:motionWhole},
\begin{equation} 
\label{eqn:motionWhole}
\begin{split}
    \argmin_{{v}_{y,1} , ..., {v}_{y,n},  {v}_{z,1} , ..., {v}_{z,n}} &=  \mathcal{L}({v}_{y,1} , ..., {v}_{y,n}, {v}_{z,1} , ..., {v}_{z,n}) \\
    &=  \mathcal{L_{\text{DPC}}} + \lambda_1 \mathcal{L}_{\text{smooth}} + \lambda_2 \mathcal{L}_{\text{DVL}}
    \end{split}
\end{equation}
where $\lambda_1$ and $\lambda_2$ are coefficients controlling the regularization strengths of the solution smoothness and fidelity to the DVL respectively.
\begin{figure*}[t]
    \includegraphics[width=0.95\linewidth]{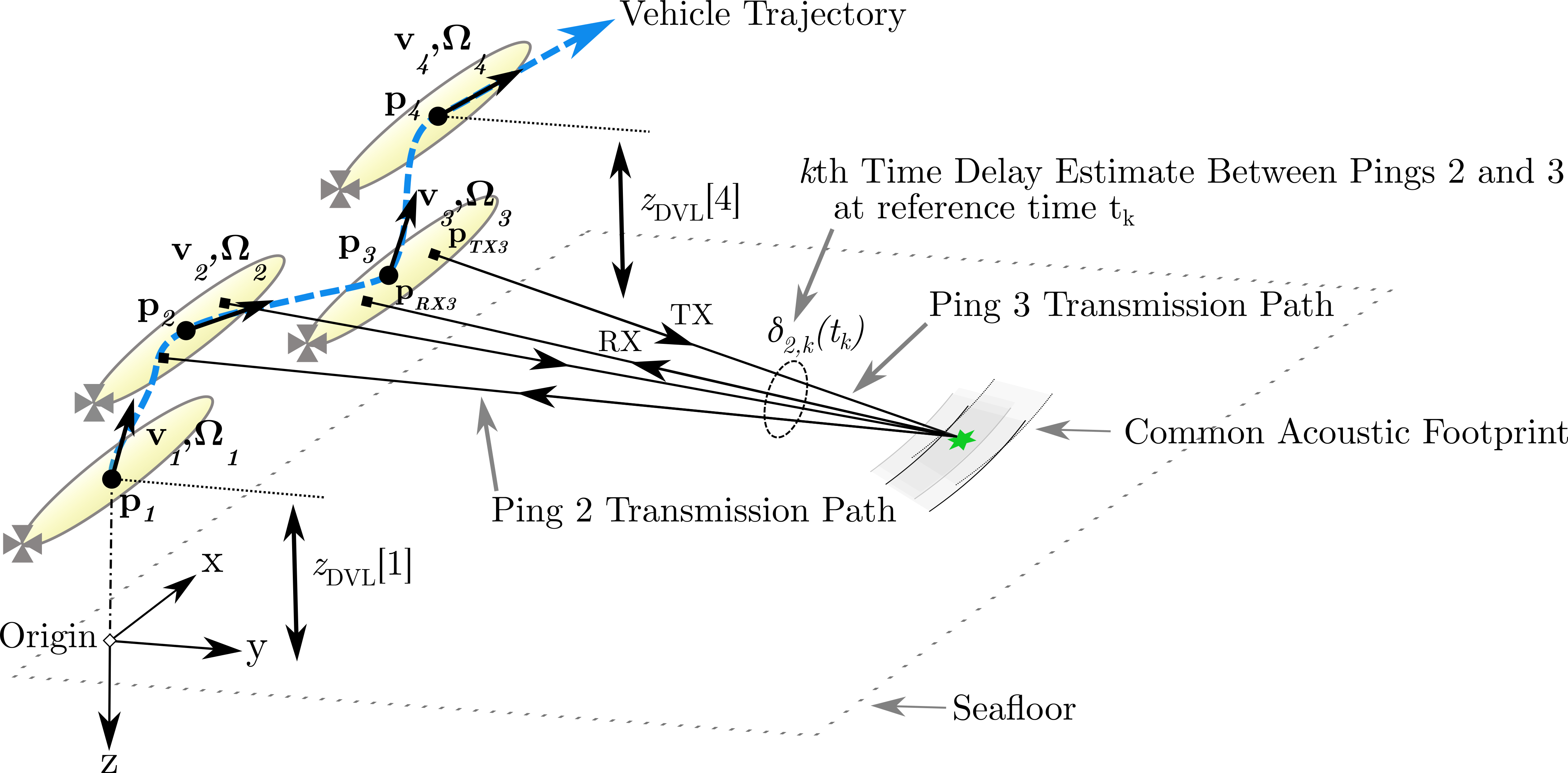}
    \centering
    \caption{Four consecutive pings from a single UUV moving across the seafloor depicting our time-of-flight model, $g$, used in the motion estimation algorithm. $\mathbf{p}_i$ is the position of ping $i$ represented as a filled dot on each vehicle, $\mathbf{v}_i$ is the velocity of ping $i$  represented by an arrow on each vehicle, and $z_{DVL}[i]$ is the altitude above the seafloor as measured by the Doppler velocity logger (DVL) on ping $i$. The origin is the coordinate system is directly below the position of the first ping.  The dotted-blue line represents the true vehicle trajectory.  Each ping's receiver and transmitter positions are determined by their initial position $\mathbf{p}_i$, velocity $\mathbf{v}_i$, and attitude / lever arms $\mathbf{\Omega}_i$. The figure shows an example time of flight pair for one common location between pings two and three.  The time-delay estimate between the time-of-flight of ping two and the seafloor and ping three and the seafloor at an associated time reference (similarly ground range also) of $t_k$ is denoted as $\delta_{2,k}(t_k)$; a dotted circle encompasses the transmissions paths compared in the time-delay estimate. The motion solution is determined by finding the $\mathbf{v_1} ... \mathbf{v_n}$ which minimizes Eq. \ref{eqn:motionWhole} given the set of time-delay estimates measured between coherent seafloor looks, $\delta_{i,k}$.}
    \label{fig:rpc_micronav_diagram}
\end{figure*}

\figurename \ref{fig:rpc_micronav_diagram} depicts the relationship of the quantities utilized in each loss term of Eq. \ref{eqn:motionWhole}.  The figure depicts four pings resulting in three overlapped ping-pairs.  We focus on pings two and three to demonstrate how a single residual $g$ is computed in Eq. \ref{eqn:motion}. The position of pings two and three results in a geometry yielding a common acoustic footprint.  The time-delay estimate between the transmission paths of ping two and ping three at reference time $t_k$ in given by $\delta_{2,k}(t_k)$.  Using the given vehicle attitude and lever arm information, ($\mathbf{\Omega}_2, \mathbf{\Omega}_3$), and the surge velocities $v_{x,2}$ and $v_{x,3}$, Eq. \ref{eqn:motionWhole} estimates $v_y, v_z$ for ping two ping and three (recall $\mathbf{v} = [v_z, v_y, v_z]$). Each ping position is represented by $\mathbf{p}_i$ in the diagram and then converted to the associated transmitter and receiver positions, $\mathbf{p}_{\text{RX}}$ and $\mathbf{p}_{\text{TX}}$, through the information provided by $\mathbf{v}_i$ and $\mathbf{\Omega}_i$. Through integration of the velocities, the absolute platform positions, $\mathbf{p}_i$, are determined relative to the origin.  The smoothness constraint described by Eq. \ref{eqn:smooth} is applied to the $y$- and $z$-velocities smoothing the final vehicle trajectory given in blue.  Finally, the DVL constraint in Eq. \ref{eqn:dvl} is applied to the $z$-components of $\mathbf{p}_i$ using the DVL measurements $z_{\text{DVL}}[i]$.

Minimization of equation Eq. \ref{eqn:motionWhole} is non-trivial to solve in closed form.  We mitigate this by using the Ceres solver library \cite{ceres-solver}. Ceres is a nonlinear least squares solver with built in auto-differentiation capability (auto-diff) coded in C++. Auto-diff provides capability to automatically and analytically evaluate the derivatives of a specified cost function, in this case equation Eq. \ref{eqn:motionWhole}.  The auto-diff functionality prevents derivation and implementation errors associated with computing derivatives when minimizing the cost function.  Auto-diff accomplishes this feat by computing derivatives via the chain rule.  Ceres provides several optimization algorithms, including gradient descent, which leverage the derivatives of the cost function (derived via auto-diff) in order to find a critical point.

To minimize equation Eq. \ref{eqn:motionWhole}, time-delay estimates between consecutive pings must be computed. We compute these estimates using the method of \cite{saebo} whereby we estimate a coarse time, $t_{coarse}$, by using quadratic interpolation of the magnitude correlation function and then refine the estimate, $t_{fine}$, by analyzing the phase of the interpolated point of the corresponding complex correlation function. The estimate of $t_{fine}$ is usually subject to phase wrapping errors due to insufficient fractional bandwidth and insufficient signal-to-noise ratio.  We overcome this hurdle by minimizing $\mathcal{L_{\text{DPC}}}$ using a two-step process utilizing a priori knowledge that $t_{coarse}$ is normally distributed around the true solution and $t_{fine}$ is normally distributed around $t_{coarse}$.  First, $\mathcal{L_{\text{DPC}}}$ is minimized using only $t_{coarse}$.  Second, we unwrap the phase around this solution and estimate $t_{fine}$. We then re-minimize  $\mathcal{L_{\text{DPC}}}$ using this unwrapped version of $t_{fine}$.  Once the time-delays are estimated and unwraped, we minimize equation Eq. \ref{eqn:motionWhole} by finding the set of ping-to-ping $v_y$ and $v_z$ velocities in ${\mathbf{v}}_1, ..., {\mathbf{v}}_{n}$ which minimize the total loss. All steps are minimized using the Levenberg–Marquardt algorithm implemented in Ceres. 

\subsection{Pre- and Post-Processing Algorithms}
The equations modeling SAS described thus far assume ideal collection conditions; this is rarely the case. The realized echoes are corrupted by a variety of sources including source-receiver spectral shaping and analog-to-digital conversion noise. The data must be conditioned for the motion estimation and image reconstruction steps to perform well.  We perform a variety of data conditioning steps prior to these operations and outline them here.

\subsubsection{Spectral Whitening}
Most sonar systems have a non-flat frequency spectrum whereby some frequencies are emphasized over others; this is undesirable when forming the synthetic aperture and can result in image artifacts during the reconstruction process. Modeling such phenomenon can be difficult so we adopt a simple data-adaptive technique called \emph{whitening} to flatten the spectrum. The spectrum is flattened by applying a gain attenuation given by Eq. \ref{eqn:whitening},

\begin{equation}
    G(f) = h \left( \frac{1}{\gamma\overline{\hat{P}(f)} + \hat{P}(f)}   \right)
    \label{eqn:whitening}
\end{equation}
where $\hat{P}(f)$ is the power estimate of frequency $f$, $\overline{\hat{P}(f)}$ is the average power over all frequencies, and $\gamma$ is a system-dependent calibration factor, and $h$ is a normalization function ensuring the minimum attenuation is $0$ dB.

\subsubsection{Time Varying Gain}
The received sonar echo is attenuated as a function of time/range by spreading and absorption losses from the medium.  We compensate for this non-linear loss of signal by applying a time varying gain (TVG) over the time-series.  Estimating an accurate TVG curve can be difficult especially if there are bright scatterers in the scene.  We mitigate this using a simple statistical method. The TVG correction is given by Eq. \ref{eqn:tvg},

\begin{equation}
    G(t) = \left( \alpha \overline{\tilde{P}(t)} + \tilde{P}(t) \right) ^ {-1}
    \label{eqn:tvg}
\end{equation}
where $\tilde{P}(t)$ is the population median power for each time sample $t$ in a batch of sonar pings, $G(t)$ is the gain correction to apply at each sample time $t$, $\overline{\tilde{P}(t)}$ is the average of the median powers at each $t$, and $\alpha$ is a system-dependent calibration factor.

\subsubsection{Dynamic Range Compression of Imagery}
SAS imagery has a dynamic range often exceeding 80dB and is beyond the range of typical displays and the human visual system in daylight.  Trivial display of SAS imagery yields an almost empty image with a few bright pixels from specular scatterers.

ASASIN produces dynamic range compressed imagery SAS suitable for human consumption and display. The algorithm used is inspired by the rational mapping function of \cite{schlick1995quantization} given by Eq. \ref{eqn:drc},
\begin{equation}
I_{\text{DRC}}[k,kk] = \frac{q I[k,kk]}{(q-1) I[k,kk] + 1}
\label{eqn:drc}
\end{equation}
where $I_{\text{DRC}}[k,kk]$ is the dynamic range compressed output at pixel location $[k,kk]$, $I[k,kk]$ is the input high dynamic range image normalized to $[0,1]$, and $q$ is a tunable brightness parameter.

Equation \ref{eqn:drc} has the advantage of having only one free parameter which is used to control the overall image brightness. For SAS imagery, the median pixel of a scene is often representative of the overall image brightness. Since Eq. \ref{eqn:drc} is bijective, it allows for a closed form solution of parameter $q$.  Given the desired brightness of the output image $I_{\text{DRC}}$, which we use the median output pixel value as a proxy for, the free parameter $q$ is computed by Eq. \ref{eqn:p},
\begin{equation}
    q = \frac{b - b  \tilde{I}}{\tilde{I} - b  \tilde{I}}
    \label{eqn:p}
\end{equation}
where $\tilde{I}$ is the median pixel value of the high dynamic range input image $I$ normalized to $[0,1]$, $q$ is the free parameter controlling image brightness, and $b\in(0,1)$ is the desired dynamic range compressed output image brightness.

\section{Experiments and Results}
\begin{figure*}[ht]
    \includegraphics[width=\linewidth]{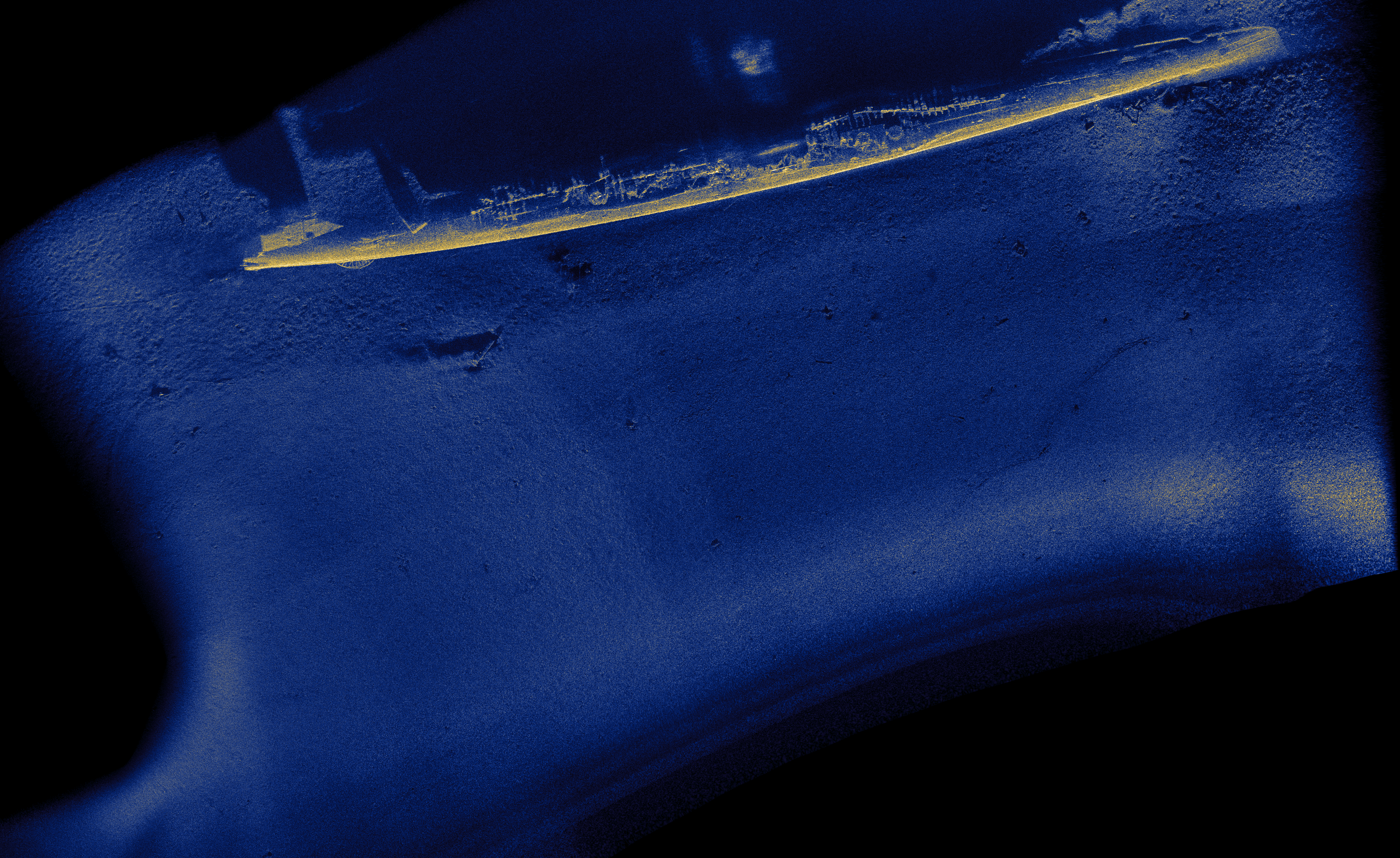}
    \caption{An example stripmap SAS image generated using ASASIN. This image was collected while the vehicle experienced significant inter-ping motion but is still able to form a well focused image due to our proposed motion estimation algorithm.}
    \label{fig:stripmap_sas}
\end{figure*}

\subsection{Stripmap Imaging}

An example image generated by ASASIN from a stripmap collection geometry is shown in \figurename \ref{fig:stripmap_sas}. The sonar flies from left-to-right along the bottom of the image (the image is collected from the port side sonar) and depicts a sunken vessel at a range far from the sonar.  This image exhibits significant inter-ping motion in the form of attitude and altitude variation but still focuses well demonstrating the efficacy of our motion estimation approach.

\subsection{Near-Field Volumetric Imaging}
ASASIN was originally implemented to generate high-resolution SAS imagery from both high-frequency ($>$100 kHz) and mid-frequency ($\sim$10 kHz) imaging sonar systems. Recently, ASASIN's image reconstruction algorithms were generalized to support generation of 3D near-field imagery of the seabed sub-bottom. The sensor used to collect this data forms a downward-looking, two-dimensional synthetic aperture from an array mounted to a surface craft \cite{brown2019simulation} that operates from 1-3 meters above the bottom. The navigation used during the image reconstruction processing is from a real-time kinematic global positioning system (RTK-GPS) aboard the craft; traditional DPC methods are not used. Adapting ASASIN to processing this sensor's data required several modifications. The primary changes were to:
\begin{enumerate}
    \item implement a bistatic ray-culling algorithm,
    \item include a sediment-water interface refraction model in determining propagation time, and
    \item enable 3D data output and provide 3D image viewers.
\end{enumerate}

First, the ray-culling algorithm shown in \figurename \ref{fig:rayculling} creates a binary mask assuming a transmit/receive pair is monostatic. This reduces both the beamforming complexity and the image reconstruction time. This monostatic approximation is valid in the standard imaging domain because the backprojection point is frequently in the far-field of the physical transmit and receive aperture. This approximation is invalid for the near-field sub-bottom sensor, and the bistatic condition must be considered.

Next, traditional SAS image formation assumes an isovelocity (constant sound speed) propagation path between the sensor and the imaging point. While this isovelocity approximation is rarely true, small deviations in sound speed have a minor effect on image quality. In the case of larger deviations, autofocusing techniques may be applied to recover some loss of focus quality \cite{callow}. Imaging within the seafloor may present the backprojection algorithm with a discontinuity in sound speed much greater than that ever observed for propagation in water. The effect of refraction must be included in the image reconstruction algorithm to create high-quality image quality. Fortunately, the backprojection algorithm is well suited to this type of modification.

Finally, the output of ASASIN was modified to produce a 3D image. This was accomplished by an iterative process where a two-dimensional output ``layer'' was generated across a range of focus depths to build up the full 3D volume. Visualization of the 3D imagery is accomplished by generating two-dimensional representations for interpretation. Planar ``slices'' through the volume are used to show a two-dimensional image within the volume. Additionally, the maximum intensity projections (MIPs) are also formed by collapsing the imagery along one of its principal axes and taking the highest intensity voxel \cite{Wallis:1991a}. For example, a slice at a depth of 11 cm is shown in \figurename \ref{fig:sliceExample}. Two targets placed in the field (solid aluminum cylinders) and two clutter object (rocks) are identified in the depth slice.

\begin{figure}[!ht]
  \centering
  \includegraphics[width=0.999\linewidth]{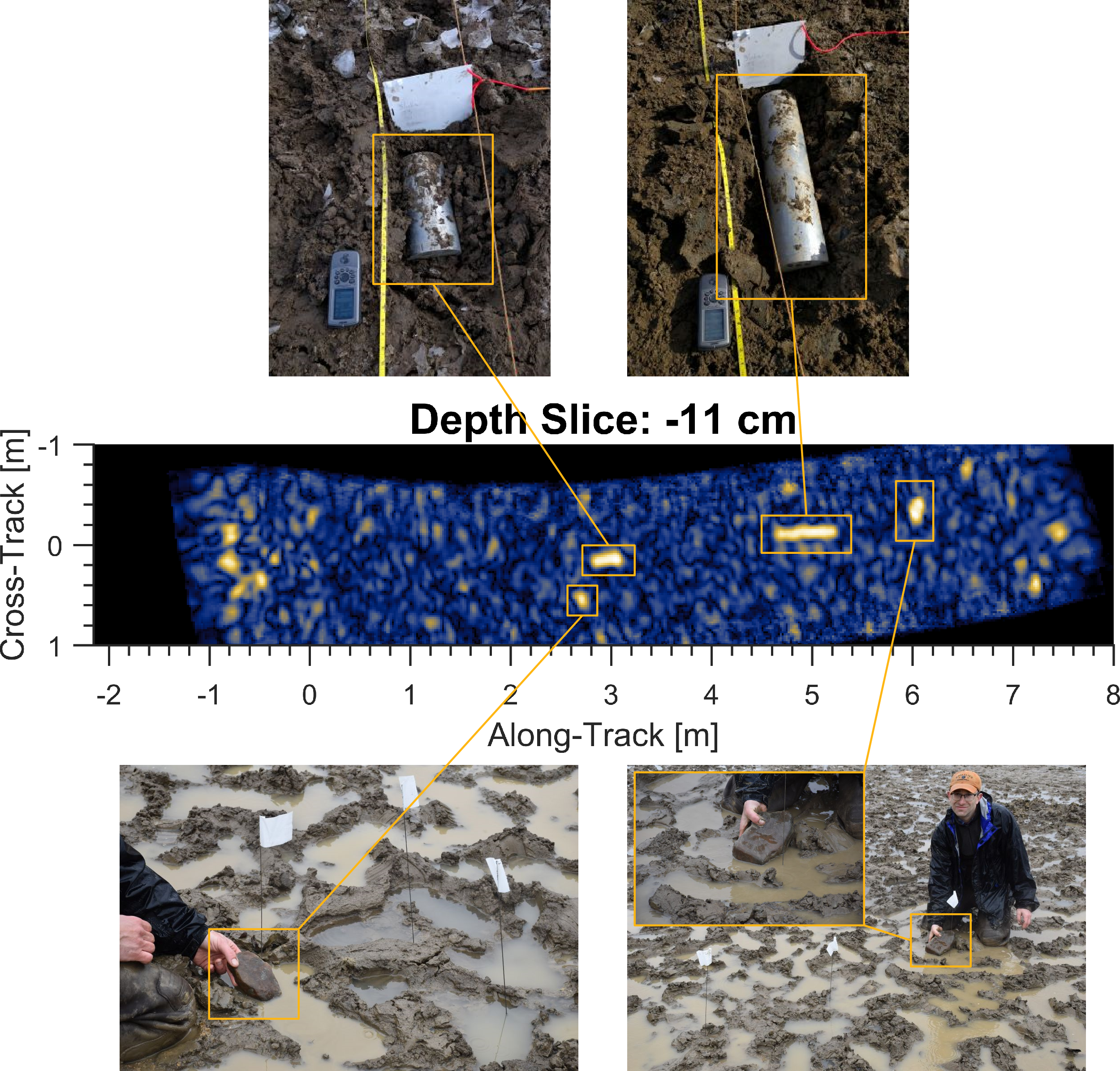}\\
  \caption{ASASIN was generalized to the problem of 3D image reconstruction for the detection of objects buried within the seabed. Here, a two-dimensional slice through a three dimensional image is shown at a depth of 11 cm into the sediment bed. A 2D synthetic aperture was used to create the middle image.}\label{fig:sliceExample}
\end{figure}

\subsection{Modeling ASASIN's Compute Performance}
\begin{figure}[!ht]
    \includegraphics[width=0.999\linewidth]{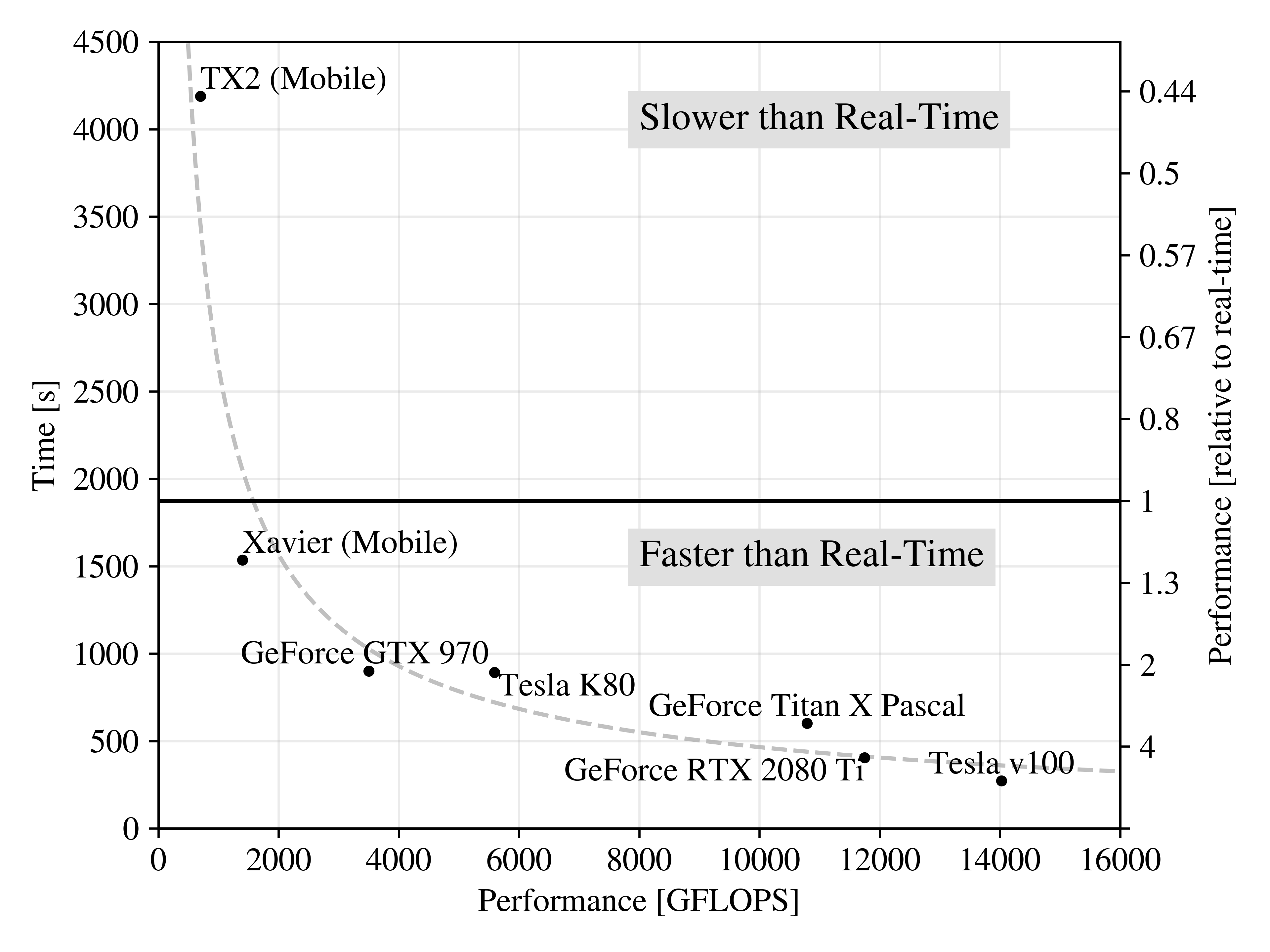}
    \centering
    \caption{ASASIN runtime performance on various GPU models; lower time is better. The thick horizontal line represents the throughput necessary to achieve real-time processing. As show, ASASIN performs faster than real-time on the NVIDIA Xavier embedded GPU.}
    \label{fig:performance}
\end{figure}

The creation of a SAS image requires computation after collection of the sensor data. The specification of these computation resources must balance processing speed, compute hardware cost, and compute hardware power requirements. In conducting post-mission analysis, one typically weighs cost against speed.  For in-situ processing embedded in a UUV, the processing speed need only occur at a real-time rate, but the power consumption should be minimized to not significantly reduce the UUV's survey duration. To support predicting the computation burden, we modeled the computational performance of ASASIN to determine the minimum hardware needed to deploy on a UUV. ASASIN is able to run on a variety of NVIDIA GPU architectures because of the flexible CUDA compiler. This affords us the ability to run the same source code on both desktop and embedded GPUs significantly reducing the amount of testing needed to reliably deploy ASASIN on board unmanned platforms.

We model the computational performance of ASASIN using a small SAS dataset representative of typical compute loads and  several GPU architectures. We process our dataset consecutively three times and report performance of the last run. This was done to ensure the memory cache was sufficiently flushed so its effects do not influence the results.  The processed data  was much larger than any available cache on the system.

\figurename \ref{fig:performance} shows the results of our performance measurements. We observe that performance versus hardware capability (reported as single-precision GFLOPS derived from \cite{list_of_nvidia_gpus}) approximately obeys a power law.  We achieve a fit shown by the gray-dashed curve in \figurename \ref{fig:performance} of $y \approx 485307 \cdot x^{-0.755}$. The thick horizontal line represents the performance needed to achieve real-time image reconstruction. The power law fit is no surprise as the image reconstruction process is highly parallelizable and its run-time inversely proportional to the number of floating point operations per second (FLOPS) available. Additionally, we measure ASASIN performance on a set of eight NVIDIA v100 GPUs contained within a single computer. In this configuration, performance scales linearly until the limit of the hard-disk throughput is reached.

\section{Conclusion}
In this work, we introduced a GPU-accelerated image reconstruction suite for SAS called ASASIN which uses a time domain backprojection image reconstruction algorithm.  We developed the design motivation of ASASIN as well as described its algorithmic components.  In particular, we gave examples demonstrating ASASIN's ability to reconstruct imagery in 2D and 3D geometries. Furthermore, we benchmarked ASASIN compute performance and developed a regression model capable of accurately predicting image reconstruction performance over a variety of GPU models for both desktop and embedded environments. Consequently, the total of our work demonstrates the feasibility of obtaining excellent image reconstruction using the time domain backprojection algorithm while simultaneously obtaining real-time performance for use on board unmanned platforms.

\section{Acknowledgments}
This research was supported in part by the U.S. Department of Defense, through the Strategic Environmental Research and Development Program (SERDP). The SERDP support was provided under the munitions response portfolio of Dr.~David Bradley. This material is based, in part, upon work supported by the Humphreys Engineer Center Support Activity under Contract No. W912HQ-16-C-0006. This work was additionally supported by the U.S. Navy. The authors would like to thank Benjamin William Thomas of the University of Bath and Thomas E. Blanford of Penn State University for their constructive remarks in improving the manuscript.


\bibliographystyle{IEEEtran}
 \bibliography{ref}

\begin{thebibliography}{10}
\providecommand{\url}[1]{#1}
\csname url@samestyle\endcsname
\providecommand{\newblock}{\relax}
\providecommand{\bibinfo}[2]{#2}
\providecommand{\BIBentrySTDinterwordspacing}{\spaceskip=0pt\relax}
\providecommand{\BIBentryALTinterwordstretchfactor}{4}
\providecommand{\BIBentryALTinterwordspacing}{\spaceskip=\fontdimen2\font plus
\BIBentryALTinterwordstretchfactor\fontdimen3\font minus
  \fontdimen4\font\relax}
\providecommand{\BIBforeignlanguage}[2]{{%
\expandafter\ifx\csname l@#1\endcsname\relax
\typeout{** WARNING: IEEEtran.bst: No hyphenation pattern has been}%
\typeout{** loaded for the language `#1'. Using the pattern for}%
\typeout{** the default language instead.}%
\else
\language=\csname l@#1\endcsname
\fi
#2}}
\providecommand{\BIBdecl}{\relax}
\BIBdecl

\bibitem{bellettini2002theoretical}
A.~Bellettini and M.~A. Pinto, ``Theoretical accuracy of synthetic aperture
  sonar micronavigation using a displaced phase-center antenna,'' \emph{IEEE
  Iournal of Oceanic Engineering}, vol.~27, no.~4, pp. 780--789, 2002.

\bibitem{brown2019simulation}
D.~C. Brown, S.~F. Johnson, I.~D. Gerg, and C.~F. Brownstead, ``Simulation and
  testing results for a sub-bottom imaging sonar,'' in \emph{Proceedings of
  Meetings on Acoustics 177ASA}, vol.~36, no.~1.\hskip 1em plus 0.5em minus
  0.4em\relax ASA, 2019, p. 070001.

\bibitem{baralli2013gpu}
F.~Baralli, M.~Couillard, J.~Ortiz, and D.~G. Caldwell, ``{GPU}-based real-time
  synthetic aperture sonar processing on-board autonomous underwater
  vehicles,'' in \emph{2013 MTS/IEEE OCEANS-Bergen}.\hskip 1em plus 0.5em minus
  0.4em\relax IEEE, 2013, pp. 1--8.

\bibitem{callow}
H.~Callow, ``Signal processing for synthetic aperture sonar image
  enhancement,'' Ph.D. dissertation, University of Canterbury, 2003.

\bibitem{cook2007synthetic}
D.~A. Cook, ``Synthetic aperture sonar motion estimation and compensation,''
  Master's thesis, Georgia Institute of Technology, 2007.

\bibitem{hunter2003simulation}
A.~J. Hunter, M.~P. Hayes, and P.~T. Gough, ``Simulation of multiple-receiver,
  broadband interferometric {SAS} imagery,'' in \emph{Oceans 2003. Celebrating
  the Past... Teaming Toward the Future (IEEE Cat. No. 03CH37492)},
  vol.~5.\hskip 1em plus 0.5em minus 0.4em\relax IEEE, 2003, pp. 2629--2634.

\bibitem{tensorflow}
``Tensorflow core guide,'' \url{https://www.tensorflow.org/guide/gpu},
  accessed: 2019-11-15.

\bibitem{marston2016volumetric}
T.~M. Marston and J.~L. Kennedy, ``Volumetric acoustic imaging via circular
  multipass aperture synthesis,'' \emph{IEEE Journal of Oceanic Engineering},
  vol.~41, no.~4, pp. 852--867, 2016.

\bibitem{ATE}
D.~C. {Brown}, I.~D. {Gerg}, and T.~E. {Blanford}, ``Interpolation kernels for
  synthetic aperture sonar along-track motion estimation,'' \emph{IEEE Journal
  of Oceanic Engineering}, pp. 1--9, 2019.

\bibitem{ceres-solver}
S.~Agarwal, K.~Mierle, and Others, ``Ceres solver,''
  \url{http://ceres-solver.org}.

\bibitem{saebo}
T.~O. Sæbø, ``Seafloor depth estimation by means of interferometric synthetic
  aperture sonar,'' Ph.D. dissertation, University of Tromsø, 2010.

\bibitem{schlick1995quantization}
C.~Schlick, ``Quantization techniques for visualization of high dynamic range
  pictures,'' in \emph{Photorealistic Rendering Techniques}.\hskip 1em plus
  0.5em minus 0.4em\relax Springer, 1995, pp. 7--20.

\bibitem{Wallis:1991a}
J.~W. Wallis and T.~R. Miller, ``Three-dimensional display in nuclear medicine
  and radiology.'' \emph{Journal of Nuclear Medicine}, vol.~32, no.~3, pp.
  534--546, 1991.

\bibitem{list_of_nvidia_gpus}
``List of nvidia graphics processing units,''
  \url{https://en.wikipedia.org/wiki/List_of_Nvidia_graphics_processing_units},
  retrieved: 2019-11-15.

\end{thebibliography}

\end{document}